\documentclass[a4paper]{article}
\usepackage{amsmath}
\usepackage{graphicx}
\usepackage{bm}
\usepackage{float}

\title{GLASS: A General Likelihood Approximate Solution Scheme}
\author{Steven Gratton\\\texttt{stg20@cam.ac.uk}\\Kavli Institute for Cosmology Cambridge, Institute of Astronomy, 
\\University of Cambridge, Madingley Road, 
\\Cambridge, CB3 0HA. United Kingdom.}
\date{August 28, 2017}

\def \l {\ensuremath{\langle}}
\def \ll {\ensuremath{\langle\langle}}
\def \r {\ensuremath{\rangle}}
\def \rr {\ensuremath{\rangle\rangle}}
\def\({\left(}
\def\){\right)}
\def\[{\left[}
\def\]{\right]}
\def\<{\left<}
\def\>{\right>}

\newcommand{\ba}{\begin{eqnarray}}
\newcommand{\ea}{\end{eqnarray}}
\newcommand{\labeq}[1] {\label{eq:#1}}
\newcommand{\refeq}[1] {(\ref{eq:#1})}
\newcommand{\labfig}[1] {\label{fig:#1}}
\newcommand{\reffig}[1] {\ref{fig:#1}}
\newcommand{\labsec}[1] {\label{sec:#1}}
\newcommand{\refsec}[1] {\ref{sec:#1}}

\newcommand{\hatC}{\ensuremath{\hat{C}}}

\newcommand{\mby}{\ensuremath{\mathbf{y}}}
\newcommand{\mbC}{\ensuremath{\mathbf{C}}}
\newcommand{\mbhatC}{\ensuremath{\mathbf{\hatC}}}

\begin{document}
\maketitle
\begin{abstract}
We present a technique for constructing suitable
posterior probability distributions in situations for which the sampling distribution 
of the data is
not known.  This is very useful for modern scientific data analysis
in the era of ``big data", for which exact likelihoods are commonly
either unknown, computationally prohibitively expensive or inapplicable because
of systematic effects in the data.   The scheme involves implicitly computing
the changes in an approximate sampling distribution as model parameters are changed
via explicitly-computed moments of statistics constructed from the data.

\end{abstract}

\section{Introduction}

Bayesian inference~\cite{jaynes} is now commonly used and understood 
as being the correct way
to learn about models from data. Posteriors for model parameters $q$ are related
via Bayes' Theorem to the product of priors for the said parameters and the
likelihood function, which is the probability for the data $x$ given the model, considered
as a function of the model parameters:
\ba
p( q | x) = \frac{p(x|q) p (q)}{p(x)}.
\labeq{bayes} 
\ea
A difficulty presents itself in making accurate inferences if the true likelihood is not known or
is unfeasible to repeatedly calculate.  Reasons might include systematics in the data rendering
an idealized likelihood unusable, or simply computational cost.  
One might still have an idea for a good choice of statistics to represent the data.  This could be inspired say by ``robustness" to systematics, by analogy to analysis procedures for idealized cases, or empirically by investigation of simulations.  If certain quantities, such as means and variances,
 are calculable (or estimatible via simulations) as functions of the model parameters in a reasonable amount
 of time, one might hope to be able to make some plausible inferences.  Whilst in the past people have been
 able to build approximations heuristically (see e.g.\ \cite{Hamimeche:2008ai, Ade:2013kta,Mangilli:2015xya}
 in a cosmological context),
 here we present a general 
 scheme for constructing suitable likelihoods in such situations.  The scheme should be relevant specifically for cosmic
 microwave background analysis, galaxy redshift surveys and the like, but is of general applicability.  

This paper is organised as follows.  First, the method is introduced and the main result derived in Sec.~\refsec{basic}.    
Next, examples
are presented in Sec.~\refsec{exs}.  Conclusions and further work are given in Sec.~\refsec{concs}.  Appendices
\refsec{altderiv}, \refsec{cumsolve} and \refsec{consistency} discuss various technical issues regarding the approximation including its derivation, practical use and validity.

\section{Basic Procedure}
\labsec{basic}

The underlying idea is to use the principle of maximum entropy to construct the broadest (i.e.\ least presumptive) 
sampling distribution consistent with a) what one assumes and b) with what one has managed to calculate about the statistics of the samples in the
context of a model~\cite{jaynes}.  This, evaluated for the data, is used as the likelihood, which, when multiplied by the prior, gives
the approximate posterior for the model given the data as in Eq.\ \refeq{bayes}.   (See Appendix \refsec{altderiv} for a complementary motivation for our approach.)  

Say, for example, one can calculate the mean $\l x \r$ and the variance $\ll x^2 \rr$ of some statistic $x$ for a model parametrized by
a parameter $q$.  (We use $\l \ldots \r$ to denote moments and $ \ll \ldots \rr$ to denote cumulants of the indicated quantities.) 
Before even calculating anything, one might also have a ``prior" $p_0(x)$ on $x$, such as it being positive for example. One then needs to maximise the entropy,
 \ba
H(p)=-\int dx \,p(x) \ln{\frac{p(x)}{p_0(x)}},
\ea
using Lagrange multipliers to impose the desired constraints on the distribution along with normalisation of $p(x)$.  This yields
\ba
p(x)= \frac{p_0(x) \exp{\left(-\lambda_x x - \lambda_{xx} x^2 \right)}}{\int dx\, p_0(x) \exp{\left(-\lambda_x x - \lambda_{xx} x^2 \right)}}.\labeq{xsd1d}
\ea
Generally, the lagrange multipliers must be solved for numerically, repeatedly evaluating $\l x \r$ and $\ll x^2 \rr$ as a function of
$\lambda_x$ and $\lambda_{xx}$ until $\l x \r (q)$ and $\ll x^2 \rr (q)$ are
obtained. The appropriate multipliers may be denoted $\lambda_x(q)$
and $\lambda_{xx}(q)$.  Substituting these into Eq.~\refeq{xsd1d} gives
our approximate sampling distribution for $x$ for the model with
parameter $q$.
If $x$ is then found to have some value, our approximate
likelihood for $q$ is then given by evaluating Eq.~\refeq{xsd1d} for that $x$. 
By analogy with conventional notation in statistical dynamics
we denote the denominator of Eq.~\refeq{xsd1d} by $Z$, and we can introduce the
``action'' $S$ (after classical/quantum mechanics) as $-\log p(x)$:
\ba
S(x,q) = -\log p_0(x) + \lambda_x(q) x + \lambda_{xx}(q) x^2 + \log Z(\lambda (q)).
\labeq{s1d}
\ea 
Multiplying by a desired prior on the parameter $q$, one then has a suitable
approximate (unnormalized-)
posterior for $q$ in light of the data $x$, appropriate for use for inference.

In principle, this procedure is easily extensible to multi-dimensional data $x^i$, $i=1,\ldots,n$, described by
multi-dimensional model parameters $q^a$, $a=1,\ldots,m$, and to use higher moments.  The Lagrange multipliers become labelled by indices, $\lambda_i$, $\lambda_{ij}$, and 
so on and will be implicit functions of the $q^a$.  The entropy becomes a multidimensional integral, and the action is
\ba
S\(x, q \)=-\log p_0(x)+\lambda_i x^i +\lambda_{\(ij\)} x^i x^j + \lambda_{\(ijk\)} x^i x^j x^k +\cdots + \log Z(\lambda(q^a))
\labeq{s}
\ea
with summation implied over repeated indices.   Here parentheses around indices indicate their symmetrization, e.g.\ $\kappa^{\( i \right.} \tau^{\left. j \right)} \equiv \frac{1}{2} \( \kappa^i \tau^j +\kappa^j \tau^i \)$, and we take all Lagrange multipliers to be symmetric since any non-symmetric part would not contribute to \refeq{s}\footnote{Alternatively one could demand, for example, that only multipliers with non-decreasing indices $i \leq j \leq \cdots \leq k $ are potentially nonzero.}.  Now    
\ba
Z(\lambda(q^a))=\int d^n x p_0 (x) e^{-\lambda_i x^i -\lambda_{\(ij\)} x^i x^j - \lambda_{\(ijk\)} x^i x^j x^k -\,\cdots}.
\labeq{z}
\ea
One varies the Lagrange multipliers until all of the desired multi-dimensional moments are matched\footnote{We typically consider matching all moments up to a given order, but in certain circumstances one might wish to match only a subset of the moments. In a two-dimensional problem for example, with variables $x$ and $y$, one might be able to calculate all the first and second moments, but only the ``auto" cubic moments $\langle x^3\rangle$ and $\langle y^3\rangle$ and not the ``cross'' ones such as $\langle x^2 y \rangle$.   In that case only the Lagrange multipliers corresponding to considered terms should be varied and the other ones, such \ $\lambda_{\(xxy\)}$ in this example, should be ignored/taken to be zero.} and the probability for the distribution becomes
\ba
p(x | q) d^n x = e^{-S\(x,q\)} d^n x.
\labeq{xsd}
\ea
In practice, the procedure rapidly becomes difficult to
perform as the dimension increases, because of the increasing difficulty of evaluating the multidimensional 
numerical integrals required to solve explicitly for the Lagrange multipliers.  

However, within a given class of models, we can get away without having
to solve explicitly for the Lagrange multipliers as follows.  Let us introduce the vector $X$ to 
denote $(x^i, x^{\(i\right.} x^{\left. j\)}, x^{\( i\right.} x^j x^{\left.k\)}, \ldots)^\mathrm{T}$.  We may indicate
a specific component of $X$ with a superscript $I$. (Nb.\ this component could contain one, two or more powers
of the $x^i$.)  Similarly, we can introduce $\lambda$ to denote the vector $\(\lambda_i, \lambda_{\( ij \)}, \ldots\)^T$ of associated Lagrange multipliers, with components $\lambda_I$.
Now we derive a set of relations between moments from the distribution given by Eq.~\refeq{xsd}.
From the form of \refeq{xsd}, we have:
\ba
\< X^I \>  (\lambda) &=& - \frac{\partial \log Z}{\partial \lambda_I},  \labeq{xfromz} \\
\<\< X^I X^J  \>\>  (\lambda) &=&  \frac{\partial^2 \log Z}{\partial \lambda_J \partial \lambda_I}. \labeq{xxfromz}
\ea
If we have managed to find $\lambda$ such that the desired $\< X \>(q)$ are obtained in a neighbourhood of $q$, then we can consider
differentiating Eq.~\refeq{xfromz} with respect to $q^a$ (sometimes denoting $\partial / \partial q^a$ by the shorthand ${}_{,a}$):
\ba
\< X^I \>_{,a} &=&- \frac{\partial^2 \log Z}{\partial \lambda_J \partial \lambda_I} \frac{\partial \lambda_J}{\partial q^a} \nonumber \\
&=& - \<\< X^I X^J  \>\> {\lambda_J}_{,a}. \labeq{xa}
\ea
Meanwhile, differentiating the action
\refeq{s} with respect to $q^a$ gives
\ba
S_{,a}=\( X^I -\< X^I \> \){\lambda_I}_{,a} \labeq{sa1}
\ea
with the $\< X^I \>$ term coming from the $\log Z$ via Eq.~\refeq{xfromz}. It is worthwhile noting that the ``prior'' $p_0(x)$ on the data has disappeared explicitly. 
Now, we should be able to invert Eq.~\refeq{xa} to solve for the ${\lambda_J}_{,a}$ in terms of the derivatives $\< X^I \>_{,a}$ of the first moments and the second cumulants $\<\< X^I X^J  \>\>$:
\ba
\lambda_{,a}=- \langle \langle X X^\mathrm{T} \rangle\rangle^{-1} \langle X \rangle_{,a} \labeq{lama}
\ea
(adopting a matrix notation).  Substituting into Eq.~\refeq{sa1} we obtain our main result:
\begin{equation}
S_{,a}=-(X- \langle X \rangle)^\mathrm{T} \langle \langle X X^\mathrm{T} \rangle\rangle^{-1} \langle X \rangle_{,a}, \labeq{sa}
\end{equation}
in which the Lagrange multipliers do not appear.  

The scheme is then to obtain the desired moments of the $X^I$, their derivatives with respect
to the $q^a$ and their second cumulants from the theory in question, ideally by calculation or potentially also by simulation.  One then uses
them in Eq.~\refeq{sa} to obtain the gradient.   This gradient is then integrated between two points in parameter space to obtain the difference in $S$ between them.   

One option for a likelihood would be to integrate  $S_{,a}$ up from a fiducial choice of $q^a$ to the values in question. (Note that as the gradient generally varies along the path and the integration takes this into account, such a likelihood would not generally be a linear expansion in parameter shifts
around a fiducial model.)  Alternatively one might integrate in steps between two nearby models under consideration in an MCMC chain for example.  When models vary smoothly with their parameters, the Romberg integration method (see e.g.\ \cite{numrec}) has been found to work well and to 
 converge quickly.  In multi-dimensional situations one can choose a path in parameter space and perform a line integral, expressing the one-dimensional 
 gradient along the path in terms of the partial derivatives of $S$ and the rate of change of parameters along the path using the chain rule. (Alternatively, one can use $S_{,a}$ directly in a sampling method that only uses the gradient of the likelihood.)

Equation \refeq{sa} is framed in terms of the means of the $X^I$, their derivatives with respect to parameters, and 
their covariance.  With the $X^I$ being powers of the $x^i$, such objects are expressible in terms of cumulants
of the latter and their derivatives with respect to parameters.  This allows one to formulate a version of  Eq.\ \refeq{sa}
``reduced down'' from the $X^I$ to the $x^i$, which may be easier to handle if the theory more directly gives cumulants of
the $x^i$ rather than of the $X^I$.  This is discussed further in Appendix \refsec{cumsolve}. 
 
 There is an approximation that is not at first sight obvious in this procedure.   Equation \refeq{sa} applies for the distribution defined by Eq.\ \refeq{xsd}.  While $\langle X \rangle$ agrees by construction between the
 underlying sampling distribution and its approximation in \refeq{xsd}, not all of the moments required in $\langle \langle X X^\mathrm{T} \rangle\rangle$ need
necessarily match.  So, to use $\langle \langle X X^\mathrm{T} \rangle\rangle$ as calculated from the underlying sampling distribution in Eq.\ \refeq{xsd} 
constitutes to \textit{using an exact result as an approximation to a term in the approximation}.  In multiple dimensions this
can lead to a breakdown in analyticity, causing the change in $S$ to become path-dependent in parameter space. 
Appendix \refsec{consistency} 
further discusses these consistency issues and potential mitigation strategies.  
In practice, one should include enough terms in $X$ to well-describe the parameter-dependent part of
the sampling distribution. Then Eq.\ \refeq{xsd} should yield a good approximate likelihood.  Reassurance in results obtained with the scheme might come through checking for stability under variation of the number of moments constrained.  The quality of the likelihood might also be judged empirically by testing its performance on suitable simulations of the data.  

\section{Examples}
\labsec{exs}

Here we examine the theory for some test cases in which the true sampling distribution 
is actually known.  The cases are motivated from cosmic microwave background (CMB) analysis,
in which one computes power spectra of spherical harmonic coefficients of (assumed-Gaussian) 
fields on the sky (see e.g.\ \cite{Ade:2015xua,Aghanim:2015xee}).

\subsection{Auto Power Spectrum Example} 

Imagine one has $2l+1$ independent Gaussianly-distributed variables $y^i$ with zero mean and the 
same variance $C$ that we are wishing to learn about.  Then the sampling distribution for the $y^i$ is just a Gaussian,
\ba
p(y|C) d^{2l+1} y= \frac{d^{2l+1} y}{(2\pi C)^{(2l+1)/2}} e^{-\sum_i \frac{{y^i}^2}{2C}} 
\labeq{ydist}
\ea
and we can see that a sufficient statistic
$\hat{C}$,
\ba
\hat{C} \equiv \frac{1}{2l+1} \sum_i {y^i}^2,
\ea 
exists with sampling distribution
\ba
p(\hatC | C) d\hatC \propto \frac{\hatC^{l-1/2} d\hatC}{C^{l+1/2}} e^{-\(l+1/2\) \frac{\hatC}{C}}.
\labeq{Csampdist}
\ea
The associated minus-log-likelihood, normalized to zero at its minimum, is
\ba
S_\mathrm{true}=\(l+1/2\)\(\frac{\hatC}{C}+\log\frac{C}{\hatC} -1 \), \labeq{sexact}
\ea
where we have dropped a $1/\hatC$ which will not affect the change in $S$ with respect to $C$.
Let us apply our method to this problem.  We shall use $\hatC$ for our $x$, and choose
to work to the lowest order possible, only constraining $\hatC$.    From  Eq.\ \refeq{ydist}
we can calculate the mean and variance of $\hatC$:
\ba
\l \hatC \r &=& C, \\
\ll \hatC^2 \rr &=& \frac{2}{2l+1} C^2. 
\ea 
Substituting in to Eq.\ \refeq{sa}, we have
\ba
 \frac{\partial S}{\partial C} = - \( \hatC - C \) \frac{2l+1}{2 C^2} 
\ea
and integrating this up actually reproduces the exact result \refeq{sexact}!

It is informative to repeat this exercise to the next level in approximation, considering 
constraining both $\l \hatC \r$ and $ \ll \hatC^2 \rr$, requiring knowledge of
up to the fourth cumulant of $\hatC$.  One again recovers the exact result, the additional
terms cancelling in the formula for $ \partial S / \partial C$.

The form \refeq{Csampdist} of the sampling distribution for $\hatC$ is linear in
$\hatC$ in the exponent. Hence it is expressible exactly with only a finite number of terms
in the form \refeq{s} that the scheme assumes, explaining why the procedure does so well in
this case.  

\subsection{Correlated Power Spectra Example}

The next case to consider is when \refeq{ydist} is generalised for $y$ to become a vector
of components, between which cross-correlations may be present.  Then the sampling distribution
becomes
\ba
p(\mby|\mbC) d^{2l+1} \mby = 
\frac{d^{2l+1} \mby}{\left| 2\pi \mbC \right|^{(2l+1)/2}} 
e^{-\frac{1}{2} \sum_i \mby_i^\mathrm{T} \mbC^{-1} \mby_i },
\ea
with \mbC\ being the covariance matrix for the components of \mby. The components of the power
spectra:
\ba
\mbhatC \equiv \frac{1}{2l+1} \sum_i \mby_i \mby_i^\mathrm{T}
\ea
are again seen to be sufficient statistics for inferences about \mbC.  

Again, from knowledge of $\l \hatC_{ij} \r$ and $ \ll \hatC_{ij} \hatC_{kl} \rr$ as a function of the model 
our procedure recovers the true form of the likelihood:
\ba
S_\mathrm{true}=\(l+1/2\)\(\text{tr} \, \mbC^{-1} \mbhatC + \log\frac{\left| \mbC \right|}{\left| \mbhatC \right|} -1 \). \labeq{scorrexact}
\ea
Again, we only need the linear constraint; repeating the procedure to quadratic order yields the same result.
One can check explicitly that the moments of \mbhatC\ satisfy Eq.\ \refeq{comm1} of Appendix~\refsec{consistency}, as needed to get both the first and second moments exactly right with only the linear constraint.

\subsection{Cross-Spectrum Example}

\begin{figure}[t]
\centering
\includegraphics[width=10cm]{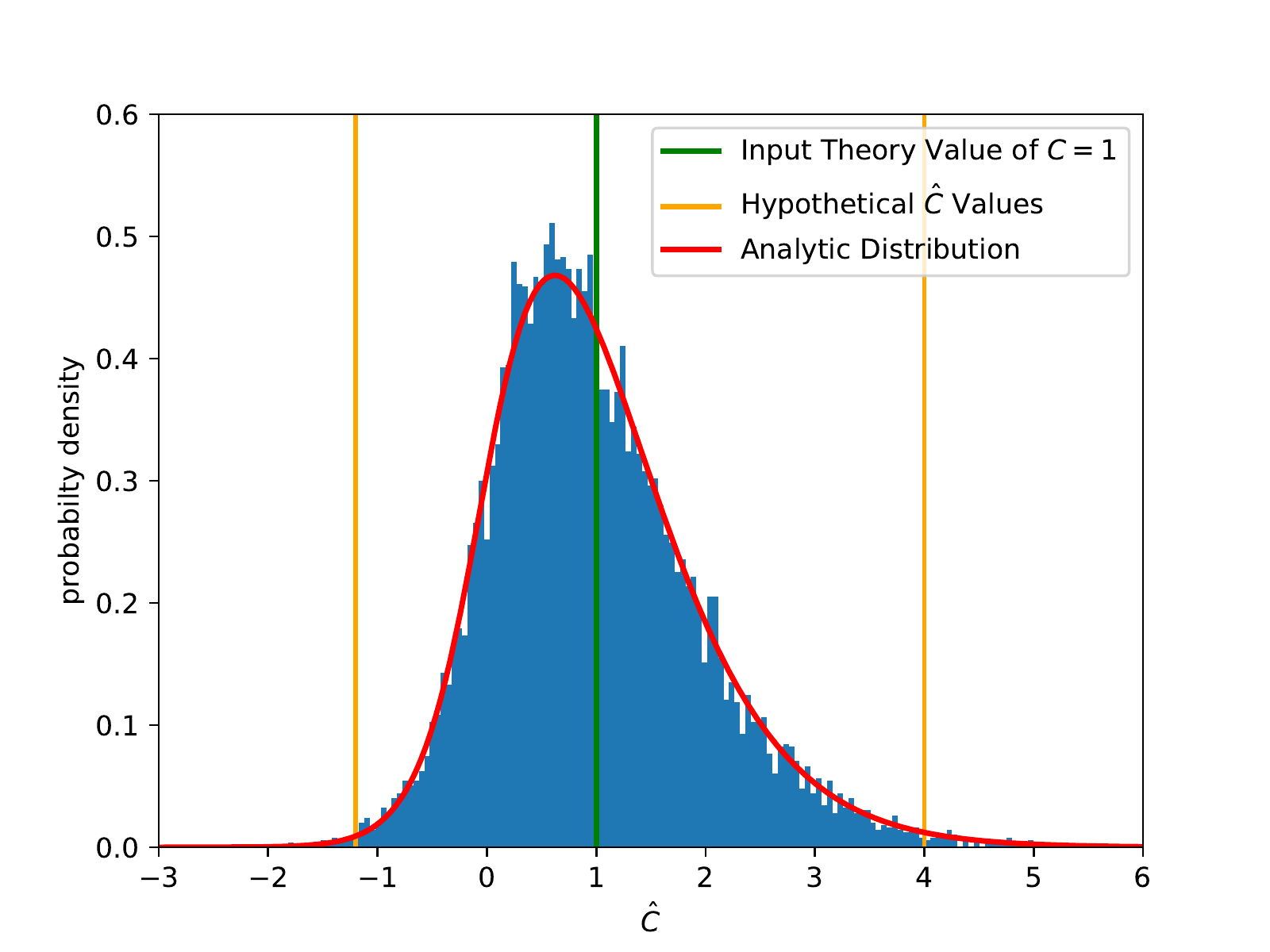}
\caption{
\labfig{distex} 
Normalized histogram of 10,000 realizations for $\hatC_{12}$, compared to analytic sampling distribution.  For this
illustration $l$ was taken to be $3$, the theory value $C=3$, and the noise levels $N_{11}=1.2$ and  $N_{22}=1.5$ were chosen.  Also shown
are three instructive data values for which the likelihood approximation will be tested.
}  
\end{figure}

Our final example is more challenging and might be considered a non-trivial test of the scheme.  Taking the case above, for a two-component vector, we may  write \mbC\ as
\ba
\mbC=
\begin{pmatrix}
C+N_{11} & C \\
C & C+N_{22}
\end{pmatrix}
\ea
and assume we are interested in making inferences about $C$.  (For example, we may imagine the two
components to be measurements of the same underlying field contaminated with independent Gaussian noise.)  We may not
know the noise levels $N_{11}$ and $N_{22}$ well enough to trust using them in a full likelihood using all the components
of \mbC.  Instead, we may try and build a likelihood using the cross-spectrum $\hatC_{12}$ alone. Such a likelihood will
hopefully be less at risk of bias in inferences about $C$.  Actually, an analytic expression for the sampling distribution
for $\hatC_{12}$ is known (see \cite{Mangilli:2015xya} for a recent use in the context of the CMB) which we can use to
compare our approximate likelihoods to.

In Fig.\ \reffig{distex} we show the distribution of $\hatC_{12}$ for 10,000 realizations and compare this to the aforementioned
analytic result.

\begin{figure}[]
\centering
\includegraphics[width=10cm]{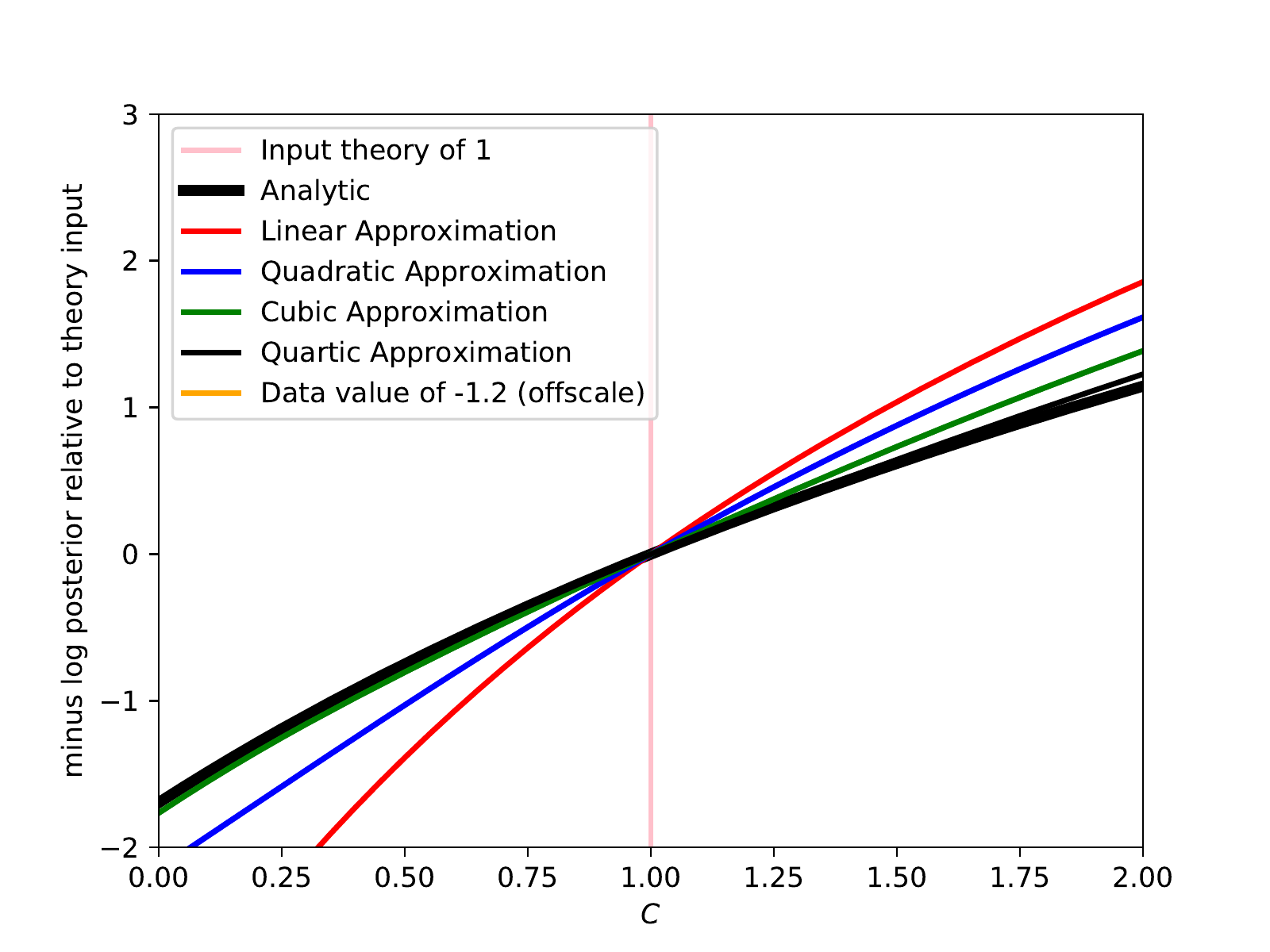}
\caption{
\labfig{postlow} 
Illustration of how the method works when the data value $\hatC=-1.2$ comes from
the low tail of the sampling distribution, using model parameters as in Fig.~\reffig{distex}.  
Different levels of approximation, coming from fitting up to the first, second, third and fourth moments,
are shown, along with the analytic result.
}  
\end{figure}

Given the Gaussianity of the \mby's, we can compute cumulants of the cross spectrum:
\ba
\l \hatC_{12} \r &=& C, \\
(2l+1) \ll \hatC_{12}^2 \rr &=& C^2+(C+N_{11})(C+N_{22}), \\
(2l+1)^2 \ll \hatC_{12}^3 \rr &=& 2C^3+6C(C+N_{11})(C+N_{22}), \\
(2l+1)^3 \ll \hatC_{12}^4 \rr &=& 6\( C^4+(C+N_{11})^2(C+N_{22})^2\right. \nonumber \\
&&\left.+6 C^2 (C+N_{11})(C+N_{22}) \,\)
\ea
and so on.  Using such cumulants we can numerically integrate up \refeq{sa} for a selection of degrees
of approximation (linear to quartic, requiring from up to quadratic to up to 8th order cumulants) and for a variety of instructive data ``realizations".   Indeed, one does well to remember that some data point could be well into the tail of the sampling distribution, particularly for
multi-dimensional data.  Therefore it is important to check the validity of a likelihood approximation for reasonable
models when some of the data is rare.  Shown in Figs.\ \reffig{postlow}, \reffig{postmid} and \reffig{posthigh} are posteriors
for $C$ (assuming a uniform prior on $C$) for $\hatC=-1.2,0.8$ and $4.0$ respectively.  
It is interesting to note
how well the approximations work, even for very rare data values.  For the low tail value, the basic linear approximation
behaves qualitatively correctly for plausible models, and as the degree of the approximation increases, the approximation
approaches the true posterior.  For the high tail value and particularly the middle value, even the linear approximation works
relatively well.
\begin{figure}[H]
\centering
\includegraphics[width=10cm]{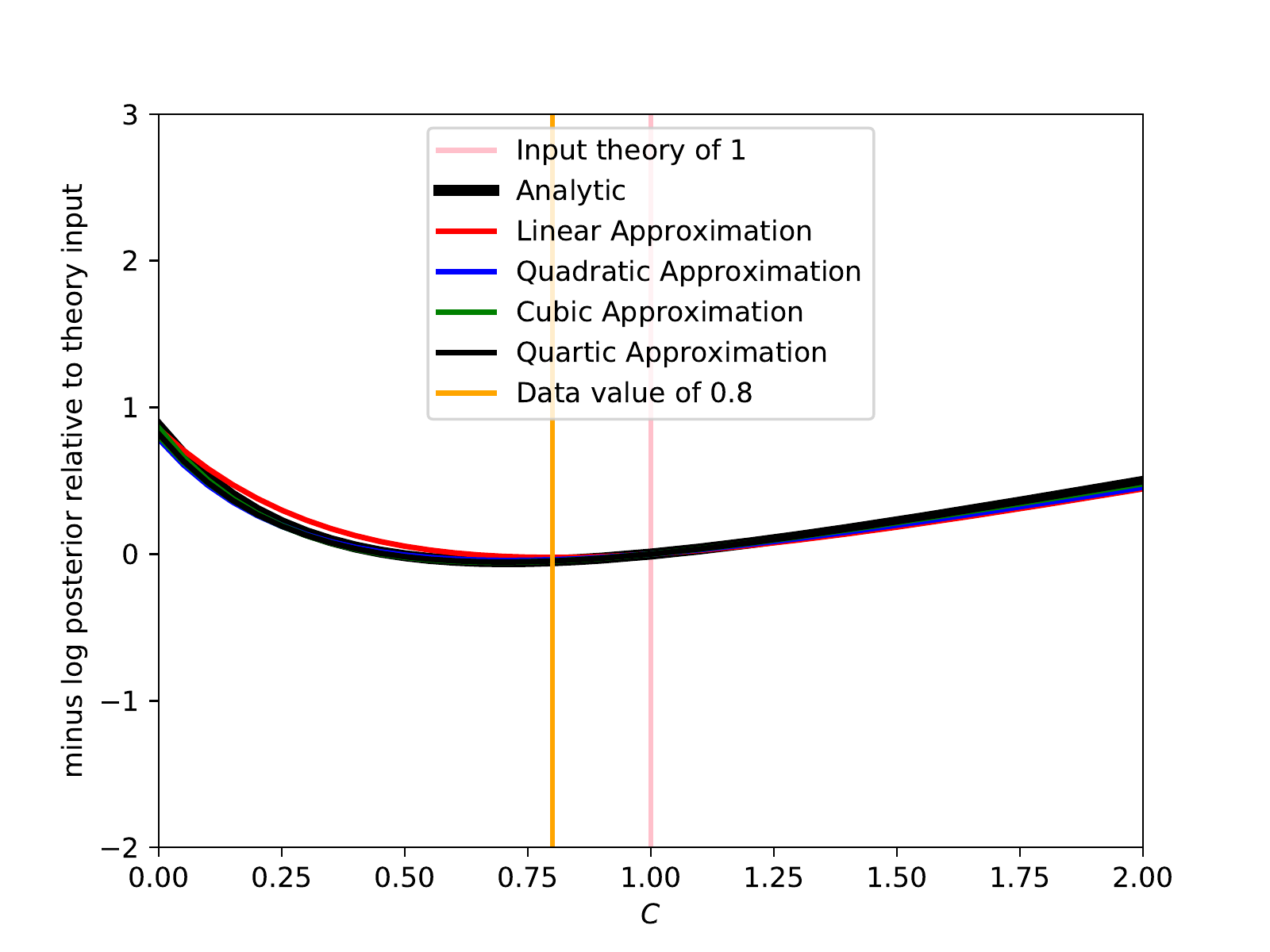}
\caption{
\labfig{postmid} 
As for Fig.~\reffig{postlow} but when the data value $\hatC=0.8$ comes from
the middle of the sampling distribution, close to the underlying model value of $C=1$.}  
\end{figure}

\begin{figure}[H]
\centering
\includegraphics[width=10cm]{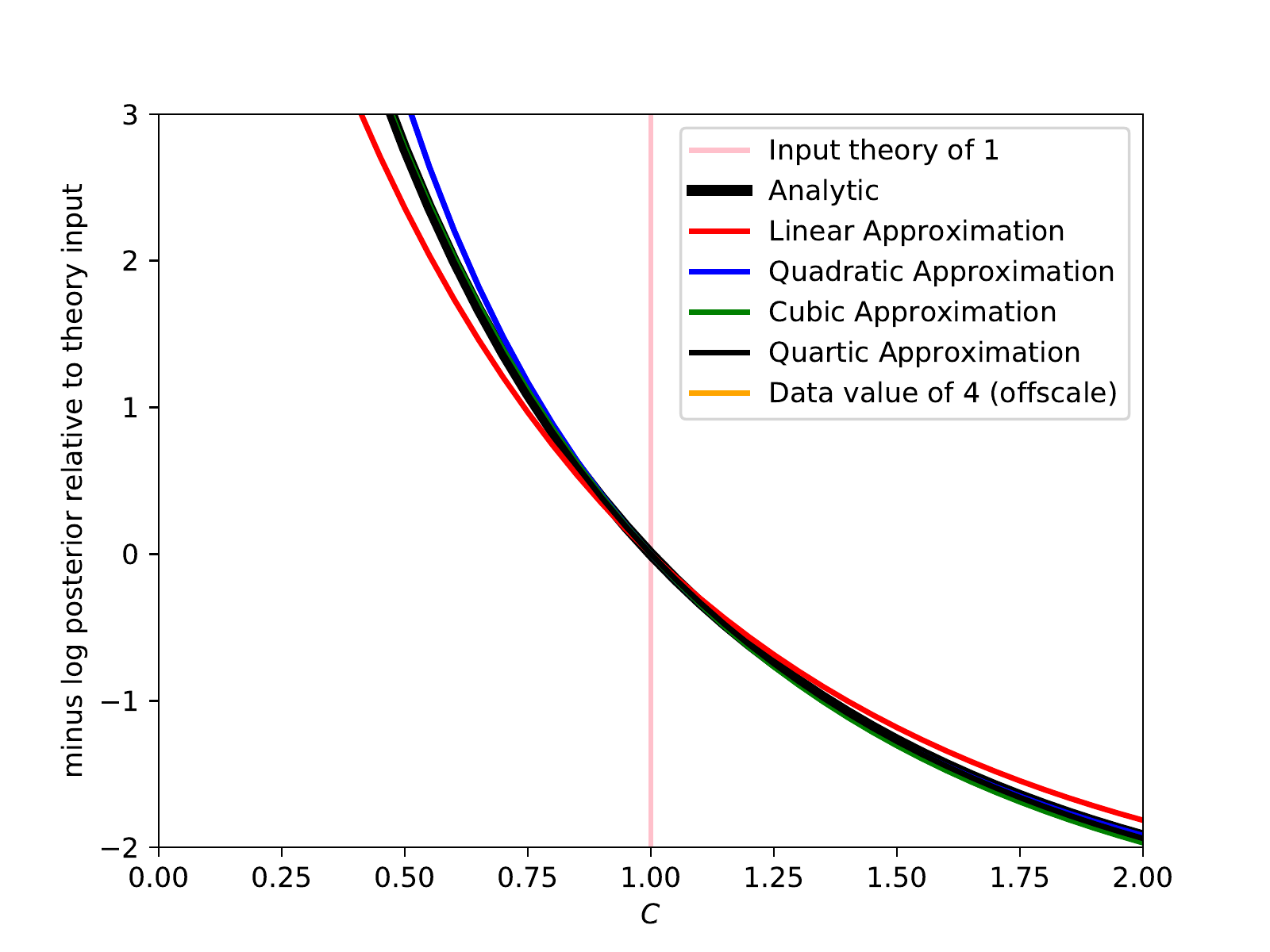}
\caption{
\labfig{posthigh} 
As for Fig.~\reffig{postlow} but when the data value $\hatC=4$ comes from
the high tail of the sampling distribution.
}  
\end{figure}

\section{Conclusion and Further Work}
\labsec{concs}

The technique presented here has some particular strengths:
\begin{description}
\item[Theoretically-underpinned] The principle of maximum entropy ensures that the
procedure uses the information it is given and makes minimal assumptions beyond that. 
\item[Calculation-based] The approximation nowhere requires the use of simulations, rather it requires
the calculation of cumulants (though one could indeed numerically estimate some of these, assuming a sufficient number of simulations 
are available, to use in the scheme if desired).
\item[Extensible] By looking for any change in the distribution as one adds in further constraints, one
can build up a feel of when the approximation is ``good enough''.  (Tests of the scheme against a limited number of realistic simulations can empirically build confidence in the approximation also.)
\end{description}

For multi-dimensional problems, it would be useful to understand the
error in the log-likelihood approximation in more detail coming from the 
potential path-dependence of the result in the parameters plane.  If this error could
be estimated to be small it might be possible then, for example, to safely use a linear
approximation instead of a quadratic one (even though the argument given in Appendix \refsec{consistency}
suggests that the latter should be more generally applicable).
A complementary step would be to develop options for manipulating higher cumulants
in order to improve the analyticity of the approximation.

Applications to CMB analysis with multi-dimensional data and tests against simulations 
will be presented in \cite{eg2017}.

\section*{Acknowledgements}

I thank Anthony Challinor, George Efstathiou and Antony Lewis
for many helpful comments and discussions over the development of this work, and Terry Iles,
Barry Nix and Andrew Pontzen for useful comments on a draft version of this paper.

\appendix

\section{Complementary Motivation for Matching Moments}
\labsec{altderiv}

Section~\refsec{basic} derives an approximate sampling distribution using maximum entropy, using
Lagrange multipliers to enforce the matching of the moments of this approximate distribution with
those calculated for the underlying one.  Here we present a complementary motivation for matching
moments.  The starting point is the Kullback-Leiber divergence 
\ba
D_\mathrm{KL}(p,q)=-\int d^n x \, p(x) \log \frac{q(x)}{p(x)}
\labeq{kl}
\ea 
which quantifies how different the probability distribution $q(x)$ is from $p(x)$.   $D_ \mathrm{KL}(p,q)$ can be thought of as the mean of the difference
in minus log probability between the approximation $q(x)$ and the true $p(x)$ with the average taken
over $p(x)$, and is minimized for $q(x)=p(x)$ (see also the discussion in \cite{2016arXiv161009018L}).  Now, imagine we wish to approximate $p(x)$ with a form
for $q(x)$ compatible with Eq.~\refeq{s}, i.e.\
\ba
q(x)=p_0(x) e^{-\alpha-\lambda_i x^i - \lambda_{ij} x^i x^j -\,\cdots}
\labeq{papp}
\ea
with a finite polynomial in the $x^i$ in the exponent.  Here we think of $\alpha$ and the $\lambda_i$, $\lambda_{ij},\,$\ldots \, as parameters that we shall vary to minimize $ D_\mathrm{KL}(p,q) $ subject
to the constraint that $q(x)$ is normalized.  Substituting into \refeq{kl}, we have:
\ba
D_\mathrm{KL}(p,q) = \int d^n x \, p(x) \(\alpha + \lambda_i x^i +  \lambda_{ij} x^i x^j + \cdots+\log \frac{p(x)}{p_0(x)}\) .
\labeq{pappsub}
\ea
We can impose the normalization constraint with a Lagrange multiplier $\beta$.  Varying with respect to $\alpha$
\ba
\frac{\partial}{\partial \alpha} \( D_\mathrm{KL}(p,q)-\beta \int d^n x \, q(x) \) &=& 0 \\
\implies \int d^n x \, p(x)- \beta \int d^n x \, q(x) &=& 0 
\ea
shows we must take $\beta=1$. Varying with respect to $\lambda_{i\ldots j}$ and substituting in $\beta=1$ then tells us that
\ba
\int d^n x \, p(x) x^i \ldots x^j - \int d^n x \, q(x)  x^i \ldots x^j.
\ea
Thus after minimization the moments of the $x^i$ that appear in the exponent in \refeq{papp} computed for the approximate distribution must match those computed for the underlying distribution.  (Note that it is not
necessary for the underlying distribution to be explicitly given, only that its appropriate moments be known.)  

So, finding the broadest probability distribution consistent with constraints on certain moments yields the 
same distribution as that coming from minimizing the Kullback-Leibler divergence of the associated functional form from the
unknown underlying distribution.

\section{Solving for the Lagrange Multiplier Derivatives}
\labsec{cumsolve}

Moments/cumulants of the $X$ and their derivatives are derivable from moments/cumulants of the $x$ and their
derivatives.  Indeed, cumulants of the $x$ are typically the things that are most straightforwardly obtainable 
from parametric models (or simulations).  Hence it is useful to be able to relate $X$-based quantities to $x$-based
ones.
By inspection of Eq.~\refeq{xsd}, we see that
\ba
\frac{\partial Z}{\partial \lambda_{ij}}&=&-\frac{\partial^2 Z}{\partial \lambda_{i} \partial \lambda_{j}},\\
\frac{\partial Z}{\partial \lambda_{ijk}}&=&\frac{\partial^3 Z}{\partial \lambda_{i} \partial \lambda_{j} \partial \lambda_{k}},
\ea
or generally:
\ba
-\frac{\partial}{\partial \lambda_{i\ldots j}} \ldots -\frac{\partial}{\partial \lambda_{p\ldots q}} Z &=&
-\frac{\partial}{\partial \lambda_i}\ldots -\frac{\partial}{\partial \lambda_j}
\ldots
-\frac{\partial}{\partial \lambda_p}\ldots -\frac{\partial}{\partial \lambda_q} Z. \labeq{Xxgeneral}
\ea
This allows us to obtain general moments of the $X$ in terms of the moments of the $x$, giving
us a straightforward route to obtaining the cumulants of the $X$ needed for Eq.~\refeq{sa}.

The relation \refeq{Xxgeneral} above suggests an alternative way of getting at the derivatives of the Lagrange
multipliers with respect to the parameters: we may start with the cumulants of the $x$, and then differentiate
them with respect to the parameters $q$.  For example, with $x$ distributed according to Eq.~\refeq{xsd}, one
has
\ba
\frac{\partial \ll x^i x^j \rr}{\partial q^a} &=& 
\frac{\partial}{\partial \lambda_r} \( \frac{\partial^2 \log Z}{\partial \lambda_j \partial \lambda_i} \) \frac{\partial \lambda_r}{\partial q^a} +\nonumber \\
&&\frac{\partial}{\partial \lambda_{rs}} \( \frac{\partial^2 \log Z}{\partial \lambda_j \partial \lambda_i} \) \frac{\partial \lambda_{rs}}{\partial q^a} +\nonumber \\
&&\dots \, .
\ea
Using commutativity of partial derivatives, the $\log Z$ derivative in the second term for example may then be rewritten as:
\ba
\frac{\partial^2 }{\partial \lambda_j \partial \lambda_i} \frac{\partial \log Z}{\partial \lambda_{rs}} 
\ea
and then we may use
\ba
\frac{\partial \log Z}{\partial \lambda_{rs}} &=& \frac{1}{Z}\frac{\partial Z}{\partial \lambda_{rs}} =
-\frac{1}{Z}\frac{\partial^2 Z}{\partial \lambda_{s} \partial \lambda_{r}} \\
&=& -\frac{1}{Z}\frac{\partial}{\partial \lambda_{s}} \(Z \frac{\partial \log Z} {\partial \lambda_{r}} \) \\&=&
-\frac{\partial \log Z} {\partial \lambda_{s}} \frac{\partial \log Z} {\partial \lambda_{r}}-
\frac{\partial^2 \log Z} {\partial \lambda_{s} \partial \lambda_{r}} 
\ea
to express the coefficient of $\frac{\partial \lambda_{rs}}{\partial q^a}$ in terms of cumulants of $x$.
With $\kappa^{i\ldots j} $ denoting $\ll x^i \ldots x^j \rr$ derived
from a distribution of the form in Eq.~\refeq{xsd}, we have:
\ba
\begin{pmatrix}
\kappa^i \\
\kappa^{\(ij\)} \\
\vdots
\end{pmatrix}_{,a}
= -
\begin{pmatrix}
\kappa^{ip} & \kappa^{ipq}+2 \kappa^{i\(p\right.} \kappa^{q\left.\)} & \cdots \\
\kappa^{ijp} & \kappa^{ijpq}+2 \kappa^{ij \(\right.p} \kappa^{q\left.\)}+ 2\kappa^{i\(\right.p} \kappa^{q\left.\)j} & \cdots \\
\vdots & \vdots &\ddots
\end{pmatrix}
\cdot
\begin{pmatrix}
\lambda_p \\
\lambda_{\(pq\)} \\
\vdots
\end{pmatrix}_{,a}
. \labeq{meq}
\ea
If we let $\kappa$ denote the vector $\(\kappa^i, \kappa^{\( ij \)}, \ldots\)^T$ of cumulants of the $x$ and
$\lambda$ denote the vector $\(\lambda_p, \lambda_{\( pq \)}, \ldots\)^T$ of Lagrange multipliers, then we may write
\ba
\kappa_{,a}= - M \lambda_{,a} \labeq{meq2}
\ea
defining $M$ to be the big matrix on the right hand side of Eq.~\refeq{meq}.  Assuming $M$ is invertible, we thus have
\ba
 \lambda_{,a}=-M^{-1} \kappa_{,a} \labeq{lama2},
 \ea
 to be compared with Eq.~\refeq{lama}.

\subsection{Cumulants as Parameters}

It is often natural to choose some of the constrained cumulants as the parameters themselves.  For example, one might
imagine that some underlying theory determines all of the cumulants of the $x^i$ in terms of a small set of parameters.  One might
wish to compare different underlying models with different fundamental parameterizations against the same data set.  In this
case one might first construct a generic likelihood in which the cumulants of the $x^i$ are set directly.  For example, if one uses
unbiased estimators of parameters as the $x^i$, then by construction their first moments are the parameters.   The higher cumulants
might then be functions of the same parameters. 

\subsection{Expanding around a Gaussian}

One can take the second derivative of Eq.~\refeq{lama2} to find
\ba
\lambda_{,ab}= M^{-1} M_{,b} M^{-1} \kappa_{,a} - M^{-1} \kappa_{,ab}
\ea
and hence expand the action to second order around a fiducial model.  A natural choice
for a fiducial model is a gaussian.  Then, in conjunction with the suggestions above
about using cumulants as the parameters themselves, we can compute the change in the
action to second order in the higher cumulants $\kappa^{ijk}$ and $\kappa^{ijkl}$.  The $M$ matrix
is upper-diagonal for a gaussian and its inverse can be analytically computed.  This expansion
may be compared to an Edgeworth expansion.

\section{Consistency of Approximation}
\labsec{consistency}

Eq.~\refeq{meq} allows one to begin to see for which circumstances a mooted approximation is possible or not.
Our procedure consists of setting some of the cumulants $\kappa^I$ as functions of the parameters as desired, and then 
hoping we can find a set of corresponding $\lambda$ and other cumulants such that Eq.~\refeq{meq} can be 
consistent. (There is some freedom in the cumulants corresponding to varying the ``prior'' term $p_0(x)$.)

Imagine for example we want a situation in which the dimension of the data matches the number
of model parameters, and we think that a likelihood constructed only from constraints on the means of the $x^i$ should
suffice. Then only the first (block-)column of the ``big" matrix in Eq.~\refeq{meq} is relevant:
\ba
\begin{pmatrix}
\kappa^i \\
\kappa^{\(ij\)} \\
\vdots
\end{pmatrix}_{,a}
= -
\begin{pmatrix}
\kappa^{ip} \\
\kappa^{ijp} \\
\vdots 
\end{pmatrix}
\cdot
\begin{pmatrix}
\lambda_p 
\end{pmatrix}_{,a}
. \labeq{meqmean}
\ea
Given that we can compute
covariances, the first (block-)row of Eq.~\refeq{meqmean} then allows us to solve for some putative $\lambda_{p,a}$.  
However,  the second (block-)row of Eq.~\refeq{meqmean} also needs to be satisfied, and then the third and so on.  One consistent
solution, for example, occurs when the covariance is independent of the model and cumulants higher than second order vanish.  
In one dimension, for arbitrary mean and variance, we can actually successively determine higher and higher cumulants to formally solve all rows of Eq.~\refeq{meqmean}.   

A counting
argument suggests a general solution however is impossible in dimensions greater than one.  First, note that the cumulant with $k$ factors, each one of $n$ variables, has $n(n+1)\cdots(n+k-1)/k!$ independent terms.  Then the $k$\textsuperscript{th} block-row involves $n\cdot n(n+1)\cdots(n+k-1)/k!$ numbers on the left, the $n$ possible derivatives of each of the terms of the $k$\textsuperscript{th} order cumulant.   
But the $k$\textsuperscript{th} block-row of the big matrix has only $n(n+1)\cdots(n+k)/(k+1)!$ numbers to vary, coming from the $k+1$\textsuperscript{th} order cumulant.  This is not enough (for $n>1$), being a factor of $(n+k) /(k+1)/n$ too small.  So, unless
appropriate functional relations exist between the cumulants of the model, this form of desired likelihood is unattainable.  

By a similar counting argument, allowing the approximate likelihood to involve quadratic constraints does not allow for solutions either. The second (block-)column introduces the $k+2$\textsuperscript{th} power cumulant, with its $n (n+1)\cdots (n+k+1)/(k+2)!$
numbers, into play,
\ba
\begin{pmatrix}
\kappa^i \\
\kappa^{\(ij\)} \\
\vdots
\end{pmatrix}_{,a}
= -
\begin{pmatrix}
\kappa^{ip} & \kappa^{ipq}+2 \kappa^{i\(p\right.} \kappa^{q\left.\)}  \\
\kappa^{ijp} & \kappa^{ijpq}+2 \kappa^{ij \(\right.p} \kappa^{q\left.\)}+ 2\kappa^{i\(\right.p} \kappa^{q\left.\)j}  \\
\vdots & \vdots 
\end{pmatrix}
\cdot
\begin{pmatrix}
\lambda_p \\
\lambda_{\(pq\)} \\
\end{pmatrix}_{,a}
.\labeq{meqquad}
\ea
This is a factor $(n+k)(n+k+1)/n/(k+1)/(k+2)$ relative to that needed to match the left hand side, for 
the $k$\textsuperscript{th} (block-)row.  So, for sufficiently large $k$,  $\sim\sqrt{n}$, there is not enough
freedom available in the cumulant.  Turning this around though, we might only expect difficulties to become
acute below some dimensionality up to a given order in the cumulant.  As the quadratic approximation needs
up to the fourth cumulant, one might suspect the scheme has a chance of working reasonably well for $n \geq20$.

For an alternative perspective, consider Eq.~\refeq{xa} again,
  If the right hand
side is indeed to be the derivatives of analytic functions, we need:
\ba
\( \lambda_{,a} \)_{,b}&=&\( \lambda_{,b} \)_{,a} \nonumber \\
\mathrm{i.e.~} \( \ll X X^T \rr^{-1} \l X \r_{,a} \)_{,b} &=& 
\( \ll X X^T \rr^{-1} \l X \r_{,b} \)_{,a}
\ea
or
\ba
\ll X X^T \rr_{,b} \ll X X^T \rr^{-1} \l X \r_{,a} = \ll X X^T \rr_{,a} \ll X X^T \rr^{-1} \l X \r_{,b}
\labeq{comm1}
\ea
to hold.  Some of the higher cumulants might then better be chosen in such a way as to satisfy Eq. \refeq{comm1}, rather
than to be equal to those calculated from the underlying theory.

\bibliographystyle{unsrt}
\bibliography{postpaper.bib}

\begin{thebibliography}{1}

\bibitem{jaynes}
E.~T. Jaynes.
\newblock {\em Probability Theory: The Logic of Science}.
\newblock Cambridge University Press, 2003.

\bibitem{Hamimeche:2008ai}
Samira Hamimeche and Antony Lewis.
\newblock {Likelihood Analysis of CMB Temperature and Polarization Power
  Spectra}.
\newblock {\em Phys. Rev.}, D77:103013, 2008.

\bibitem{Ade:2013kta}
P.~A.~R. Ade et~al.
\newblock {Planck 2013 results. XV. CMB power spectra and likelihood}.
\newblock {\em Astron. Astrophys.}, 571:A15, 2014.

\bibitem{Mangilli:2015xya}
A.~Mangilli, S.~Plaszczynski, and M.~Tristram.
\newblock {Large-scale cosmic microwave background temperature and polarization
  cross-spectra likelihoods}.
\newblock {\em Mon. Not. Roy. Astron. Soc.}, 453(3):3174--3189, 2015.

\bibitem{numrec}
W.H. Press, S.A. Teukolsky, W.T. Vetterling, and B.P. Flannery.
\newblock {\em Numerical Recipes: The Art of Scientific Computing}.
\newblock Cambridge University Press, 1986.

\bibitem{Ade:2015xua}
P.~A.~R. Ade et~al.
\newblock {Planck 2015 results. XIII. Cosmological parameters}.
\newblock {\em Astron. Astrophys.}, 594:A13, 2016.

\bibitem{Aghanim:2015xee}
N.~Aghanim et~al.
\newblock {Planck 2015 results. XI. CMB power spectra, likelihoods, and
  robustness of parameters}.
\newblock {\em Astron. Astrophys.}, 594:A11, 2016.

\bibitem{eg2017}
George Efstathiou and Steven Gratton.
\newblock In preparation.

\bibitem{2016arXiv161009018L}
R.~H. {Leike} and T.~A. {En{\ss}lin}.
\newblock {Optimal Belief Approximation}.
\newblock arXiv:1610.09018, 2016.

\end{thebibliography}

\end{document}